\documentclass[twocolumn]{aastex62}

\begin{document}
\title{\bf \large Evidence for Low Radiative Efficiency or Highly Obscured Growth of $z>7$ Quasars}

\author{Frederick B. Davies}
\affiliation{Department of Physics, University of California, Santa Barbara, CA 93106-9530, USA}

\author{Joseph F. Hennawi}
\affiliation{Department of Physics, University of California, Santa Barbara, CA 93106-9530, USA}
\affiliation{Max Planck Institut f\"{u}r Astronomie, K\"{o}nigstuhl 17, D-69117 Heidelberg, Germany}

\author{Anna-Christina Eilers}
\altaffiliation{IMPRS Fellow}
\affiliation{Max Planck Institut f\"{u}r Astronomie, K\"{o}nigstuhl 17, D-69117 Heidelberg, Germany}

\begin{abstract}

 The supermassive black holes (SMBHs) observed at the centers of all
 massive galaxies are believed to have grown via luminous accretion
 during quasar phases in the distant past. The fraction of inflowing
 rest mass energy emitted as light, the radiative efficiency, has been
 inferred
 to be 10\%, in agreement with expectations from thin disk accretion
 models. But the existence of billion solar-mass SMBHs powering
 quasars at $z > 7$ challenges this picture: provided they respect the
 Eddington limit, there is not enough time to grow $z>7$ SMBHs from
 stellar remnant seeds unless the radiative efficiency is below
 10\%. Here we show that one can constrain the radiative efficiencies of
 the most distant quasars known using foreground neutral intergalactic
 gas as a cosmological-scale ionizing photon counter. From the
 Ly$\alpha$ absorption profiles of ULAS J1120+0641 ($z=7.09$) and ULAS
 J1342+0928 ($z=7.54$), we determine posterior median radiative
 efficiencies of 0.08\% and 0.1\%, respectively, and the combination
 of the two measurements rule out the canonical 10\% efficiency at 99.8\%
 credibility after marginalizing over the unknown obscured fraction.
  This low radiative efficiency implies rapid mass
 accretion for the earliest SMBHs,
 greatly easing the tension between the age of the Universe
 and the SMBH masses. However, our measured efficiency may instead reflect
 nearly complete obscuration by dusty gas in the
 quasar host galaxies over the vast majority of their SMBH growth.
  Assuming 10\% efficiency during unobscured phases, 
 we find that the obscured fraction would be $>82\%$ at 95\% credibility, 
 and imply a
  $25.7^{+49.6}_{-16.5}$ times larger obscured than unobscured luminous quasar population at $z>7$.

\end{abstract}

\section{Introduction}

The quasar phenomenon has been studied for more than 50 years \citep{Schmidt63}. 
Quasar central engines are believed to be
accretion disks feeding material onto supermassive black
holes \citep{Rees84}.
 In the standard picture of SMBH growth,
the rest mass energy of accreted mass is divided between radiation
and black hole mass growth via the radiative efficiency $\epsilon$,
implying that emission of quasar light is concomitant with black hole growth. 
In the local Universe, dormant
supermassive black holes reside at the centers of all massive
galaxies, and galaxies with more stellar mass in a spheroidal bulge
host more massive black holes \citep{FM00,Gebhardt00}.
The connection between distant quasar ``progenitors'' and ``relic'' supermassive black holes is
encapsulated in the Soltan argument, which states that the
integrated emission from quasars over cosmic time is proportional
to the total mass in supermassive black holes today via the radiative efficiency \citep{Soltan82},
\begin{equation} \label{eqn:soltan}
\rho_{\rm BH}(z=0) = \int_0^\infty \frac{dt}{dz} dz \int_{L_{\rm min}}^\infty \frac{1-\epsilon}{\epsilon} \frac{L_{\rm q}}{c^2} \Phi(L,z) dL,
\end{equation}
where $\Phi(L,z)$ is the quasar luminosity function in some observed band and $L_{\rm q}$ represents the bolometric luminosity of a quasar with observed luminosity $L$.
It then follows that the radiative efficiency can be inferred
by 
commensurating the energy density of quasar light with
the inferred mass density of supermassive black holes in the
local Universe.
Applications of this argument by various groups
have measured radiative efficiencies of $\approx10\%$ (e.g. \citealt{YT02,Shankar09,Ueda14})
after statistically correcting for the obscured quasar population,
consistent with predictions of analytic thin disk
accretion models
in general relativity (e.g. \citealt{Thorne74}).  
However, current understanding is that these thin disk models represent an idealization
as they fail to reproduce quasar spectral energy distributions (e.g. \citealt{KB99}). 
Numerical simulations of quasar accretion disks reveal a more complex
picture of geometrically thick disks supported by radiation pressure,
within which a substantial fraction of the radiation can be advected into the central black
hole, potentially dramatically lowering the radiative
efficiency (e.g. \citealt{Sadowski14}). 

The bolometric luminosity of a quasar accretion disk is typically written as
\begin{equation} \label{eqn:acc}
L_{\rm bol} = \epsilon \dot{M} c^2 = \frac{\epsilon}{1-\epsilon} \dot{M}_{\rm BH} c^2,
\end{equation}
where $\dot{M}$ is the total mass inflow rate and $\dot{M_{\rm BH}}$ is the growth rate of the black hole.
The maximum luminosity of a quasar can be estimated by equating the gravitational acceleration from the black hole with radiation pressure on electrons in the infalling gas, known as the Eddington luminosity,
\begin{equation}
L_{\rm edd} = \frac{4\pi GM_{\rm BH}cm_{\rm p}}{\sigma_{\rm T}}.
\end{equation}
From equation~(\ref{eqn:acc}), the characteristic timescale for growing a black hole at the Eddington limit, the Salpeter time $t_{\rm S}$, is then
\begin{equation} \label{eqn:salp}
t_{\rm S} = \frac{1-\epsilon}{\epsilon} \frac{4\pi G m_{\rm p}}{c\sigma_{\rm T}} \approx 45\,{\rm Myr} \times \frac{(1-\epsilon)/\epsilon}{9}.
\end{equation}
Assuming a fixed Eddington ratio $L/L_{\rm Edd}$, a black hole with seed mass $M_{\rm seed}$ at time $t_{\rm seed}$ will then grow as
\begin{equation}
M_{\rm BH}(t) = M_{\rm seed} e^{(L/L_{\rm Edd})([t-t_{\rm seed}]/t_{\rm S})}.
\end{equation}
A lower radiative efficiency would decrease $t_{\rm S}$, and thus could alleviate the tension with growing
supermassive black holes at the highest redshifts.

Luminous quasars with
$\gtrsim10^9$ $M_\odot$ black holes have been discovered at $z > 7$
when the Universe was less than 800 Myr
old \citep{Mortlock11,Banados18,Yang18,Wang18}. 
Growing the observed $10^9$ $M_\odot$ black holes at $z\gtrsim7$ from a
$100$ $M_\odot$ initial seed requires $\approx16$
e-foldings, which for $\epsilon=0.1$ corresponds to continuous Eddington-limited accretion for
roughly the entire age of the Universe at that time. 
It seems implausible that these black holes have been growing since
the Big Bang. However, demanding a later formation epoch, consistent with expectations
for the death of the first stars in primordial galaxies at $z\sim 20$--$50$ (e.g. \citealt{Tegmark97}), 
implies seeds in excess of $1000$ $M_\odot$ \citep{Mazzucchelli17,Banados18} which are then inconsistent with being
stellar remnants \citep{Heger03}. 

Two classes of models have been proposed to resolve this tension. In the
first, the black holes grow faster, either by explicitly violating the Eddington luminosity limit (e.g. \citealt{VR05})
or by accreting at a much lower radiative efficiency (e.g. \citealt{Madau14}). In the 
second, the initial seeds were much more massive than stellar remnants,
either by forming monolithically via direct collapse of primordial
gas (e.g. \citealt{BL03}) or by coalescence of a dense Population III stellar cluster (e.g. \citealt{Omukai08}). 
A method to directly measure the radiative efficiency of the highest redshift
quasars would shed light on this tension and distinguish between these models.
Indeed, the radiative efficiency inferred from the Soltan argument is both
indirect and has negligible contribution from the rare $z > 7$ quasar population.

The highest redshift quasars known reside within the ``epoch of
reionization,'' when the first stars, galaxies, and accreting black
holes ionized the hydrogen and helium in the Universe for the first
time after cosmological recombination \citep{LF13,DF18}.  During reionization, abundant neutral hydrogen 
in the intergalactic medium (IGM) is expected to imprint two distinct Ly$\alpha$ absorption
features on the rest-frame ultraviolet (UV) spectra of quasars. First,
the ``proximity zone'' of enhanced  Ly$\alpha$ transmission resulting from the quasar's own ionizing radiation
will be truncated by neutral hydrogen along our line-of-sight \citep{CH00}.  Second, a damping wing
signature redward of rest-frame Ly$\alpha$ will be present, arising
from the Lorentzian wings of the Ly$\alpha$ resonant absorption
cross-section \citep{ME98}. 
The two highest redshift quasars known, ULAS
J1120+0641 \citep{Mortlock11} (henceforth J1120+0641) at $z=7.09$, and
ULAS J1342+0928 \citep{Banados18} (henceforth J1342+0928) at $z=7.54$,
both exhibit truncated proximity zones \citep{Bolton11,Davies18b}
(compared to similarly-luminous quasars at
$z\sim6$--$6.5$, \citealt{Eilers17}) and show strong evidence for damping
wing absorption \citep{Mortlock11,Bolton11,Greig17b,Banados18,Davies18b}.
As we show below, an extension of the Soltan argument to \emph{individual quasars} is uniquely
possible at $z\gtrsim7$ due to the presence of neutral hydrogen in the
IGM along our line of sight to the quasar.

\begin{figure*}[htb]
\begin{center}
\resizebox{18cm}{!}{\includegraphics[trim={1em 1em 1em 1em},clip]{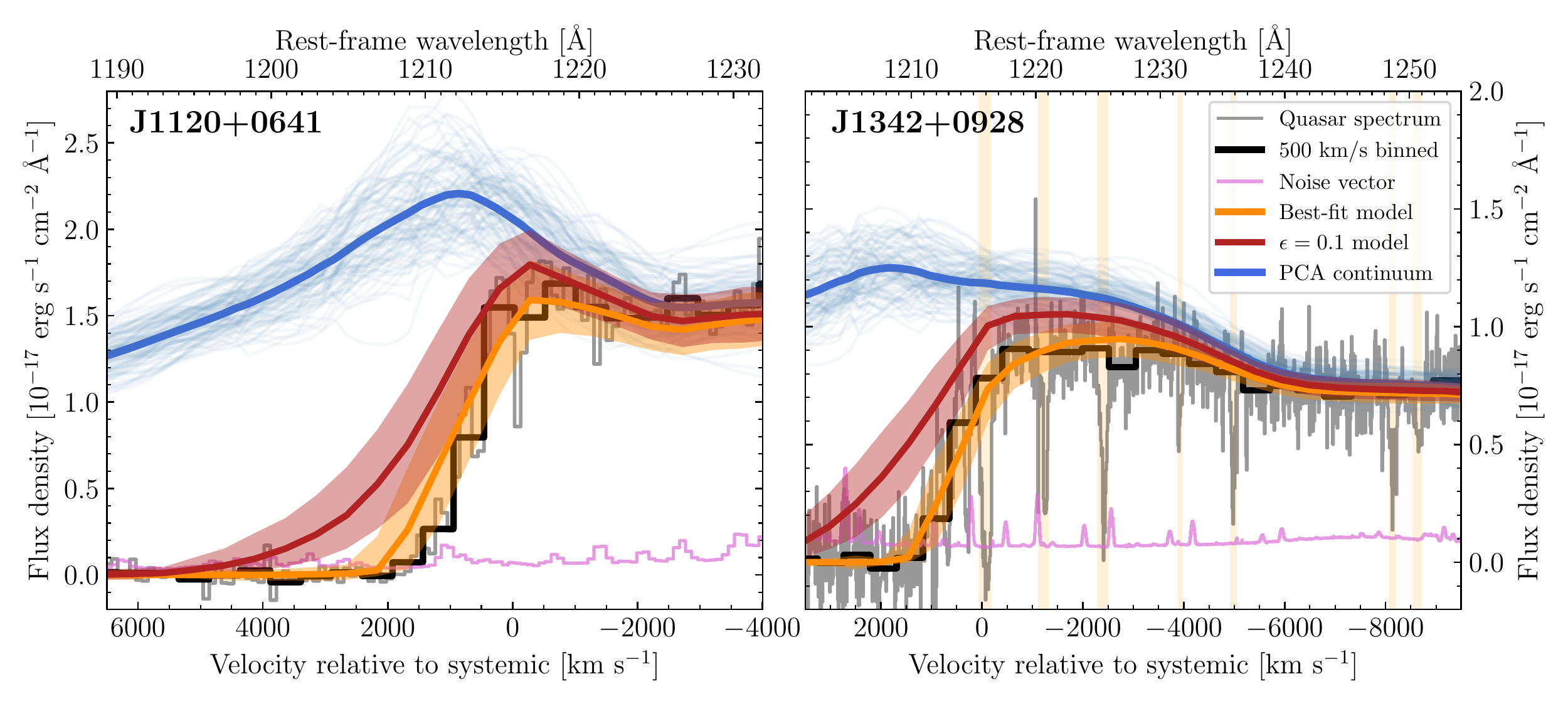}}
\end{center}
\caption{Quasar spectra and model fits close to rest-frame Ly$\alpha$. The grey and pink curves show the observed spectra and corresponding $1\sigma$ noise for J1120+0641 (left, VLT/FORS2, \citealt{Mortlock11}) and J1342+0928 (right, Magellan/FIRE, \citealt{Banados18}). The black curves show the 500 km/s binned spectra used in our statistical analysis. The vertical shaded bands correspond to foreground metal absorption systems which were masked prior to binning. The thick blue curves show the PCA models for the intrinsic quasar spectra, while the thin blue curves show 100 draws from the distribution of covariant prediction uncertainty. The orange curves show the best-fit mean absorption models, corresponding to ($\langle x_{\rm HI} \rangle$, $N_{\rm ion}$) = (0.65, $1.2\times 10^{71}$) for J1120+0641 and ($\langle x_{\rm HI} \rangle$, $N_{\rm ion}$) = (0.90, $7.0\times 10^{70}$) for J1342+0928. The red curves show models assuming a fully neutral Universe ($\langle x_{\rm HI} \rangle=1$) and $N_{\rm ion} = 5.7\times10^{72}$ and $1.8\times10^{72}$ for J1120+0641 and J1342+0928, respectively, corresponding to the maximum Ly$\alpha$ absorption with $N_{\rm ion}$ from equation~(\ref{eqn:lrsa}) assuming $\epsilon=0.1$. The shaded regions around the orange and red curves correspond to the central 68\% scatter of forward modeled mock spectra.}
\label{fig:spectra}
\end{figure*}

\begin{figure*}[htb]
\begin{center}
\resizebox{18cm}{!}{\includegraphics[trim={4em 1em 4em 1em},clip]{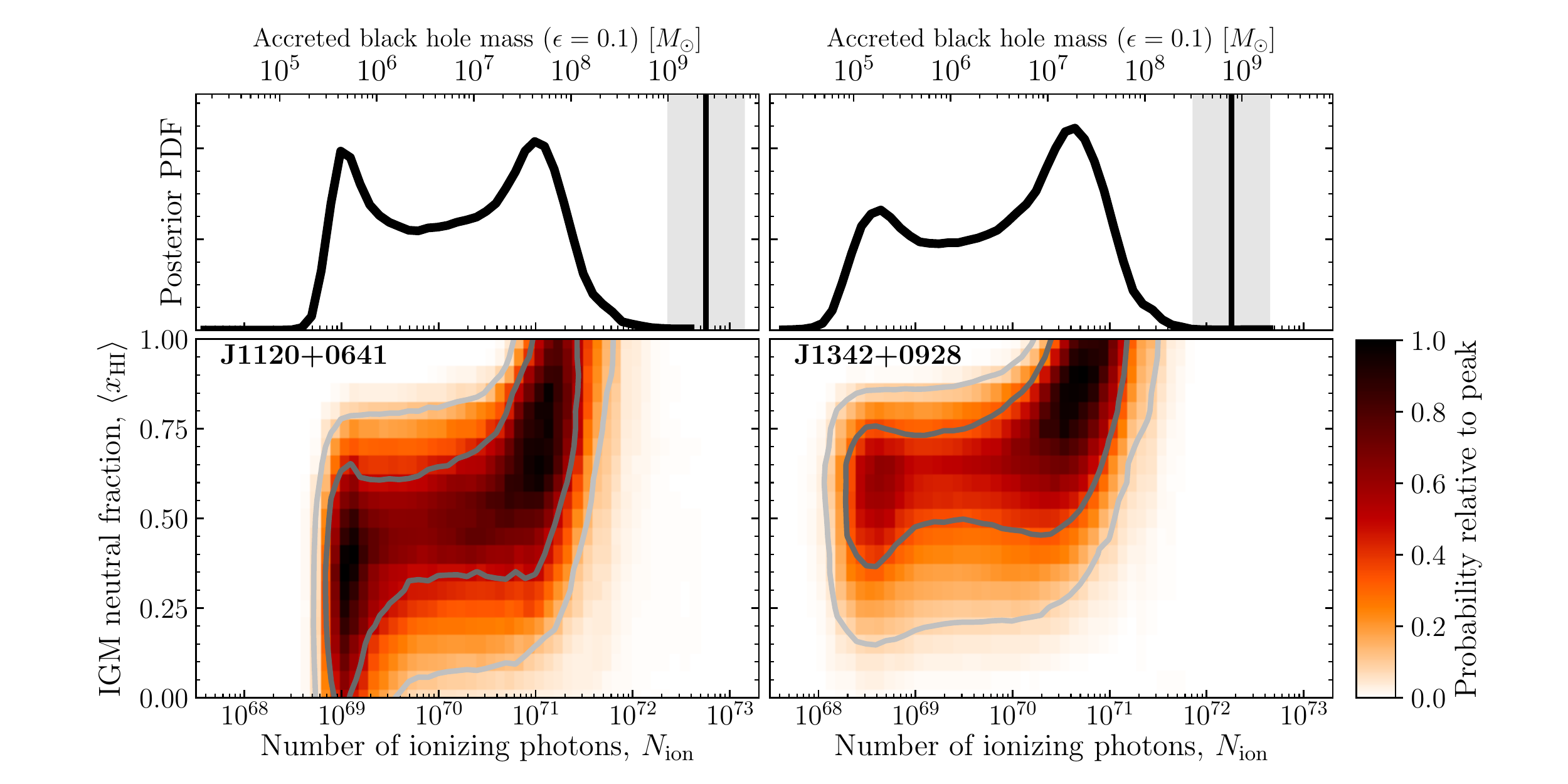}}
\end{center}
\caption{Joint posterior PDF of IGM neutral fraction and number of emitted ionizing photons. The lower panels show the two-dimensional posterior PDFs of $\langle x_{\rm HI} \rangle$ and $N_{\rm ion}$ inferred from the spectra of J1120+0641 (left) and J1342+0928 (right). The inner and outer grey contours enclose the central 68\% and 95\% probability, respectively. The top panels show the corresponding marginalized posterior PDFs for $N_{\rm ion}$. The vertical lines in the top panels indicate the expected $N_{\rm ion}$ from equation~(\ref{eqn:lrsa}), assuming $\epsilon=0.1$ and the measured black hole masses of J1120+0641 ($M_{\rm BH}=2.47\times10^9$ $M_\odot$) and J1342+0928 ($M_{\rm BH}=7.8\times10^8$ $M_\odot$), with a shaded region indicating a $1\sigma$ systematic uncertainty of 0.4 dex in the mass measurements.}
\label{fig:2dpost}
\end{figure*}

\section{The Local Reionization Soltan Argument} \label{sec:soltan}

The simple form of our analogy to the Soltan argument is as follows. 
The imprint of the neutral IGM on a reionization-epoch quasar spectrum
constrains the total number of ionizing photons that the quasar ever emitted, $N_{\rm ion}$,
which is proportional to the total accreted black hole mass, $\Delta M_{\rm BH}$, 
via the radiative efficiency $\epsilon$. 
From measurements of $N_{\rm ion}$ and $M_{\rm BH}$, we can thus constrain the
average radiative efficiency during the entire growth history of the central SMBH.
Below we explain this argument in more detail.

The total number of ionizing photons emitted by a quasar can be written as
$N_{\rm ion} = \int \dot{N}_{\rm ion}(t) dt$,
where $\dot{N}_{\rm ion}(t)$ is the quasar's ionizing photon emission rate. Assuming unobscured emission along our
line of sight, a constant bolometric correction $L_{\rm bol} = C_{\rm ion} \dot{N}_{\rm ion}$, and a constant radiative efficiency $\epsilon$, we can write
\begin{equation} \label{eqn:nion2}
N_{\rm ion} = \int \frac{L_{\rm bol}(t)}{C_{\rm ion}} dt = \frac{\epsilon}{1-\epsilon} \frac{c^2}{C_{\rm ion}} \int \dot{M}_{\rm BH}(t) dt \propto \Delta M_{\rm BH}.
\end{equation}
That is, given a bolometric correction and radiative efficiency, we can translate the number of ionizing photons into the mass growth of the black hole, $\Delta M_{\rm BH} \propto N_{\rm ion}$.

We assume the luminosity-dependent bolometric correction from $M_{1450}$ given in Table 3 of \citet{Runnoe12}\footnote{We additionally include the factor of 0.75 advocated by \citet{Runnoe12} to correct for viewing angle bias.}, and convert from $M_{1450}$ to ionizing luminosity following the \citet{Lusso15} spectral energy distribution (SED), resulting in 
$C_{\rm ion}=8.63\times10^{-11}$ erg per ionizing photon. In the following, we neglect the uncertainty in this conversion because the uncertainty in the quasar SED is substantially smaller than the systematic uncertainty in the black hole mass.
For J1120+0641 and J1342+0928, we compute ionizing photon emission rates of $\dot{N}_{\rm ion}^{\rm J1120}=1.2\times10^{57}$\,s$^{-1}$ and $\dot{N}_{\rm ion}^{\rm J1342}=1.4\times10^{57}$\,s$^{-1}$, and bolometric luminosities of $L_{\rm bol}^{\rm J1120}=2.7\times10^{13}$\,$L_\odot$ and $L_{\rm bol}^{\rm J1342}=3.1\times10^{13}$\,$L_\odot$, respectively.

Similar to analyses of the original Soltan argument, the census of ionizing photons recorded by the surrounding IGM
must be modified to account for obscured phases when ionizing photons are absorbed before escaping the quasar host.
That is, any ionizing photons emitted by the quasar which did not reach the IGM along our particular line of sight will be absent from
our accounting of $N_{\rm ion}$ from the spectrum.
Accordingly, we predict that black hole mass, radiative efficiency, and the number of ionizing photons should obey
\begin{equation} \label{eqn:lrsa}
  \Delta M_{\rm BH} = 10^9\,M_\odot \times \left(1-f_{\rm obsc}\right)^{-1} \left(\frac{(1-\epsilon)/\epsilon}{9}\right) \left(\frac{N_{\rm ion}}{2.3\times10^{72}}\right), 
\end{equation}
where $f_{\rm obsc}$ is the fraction of emitted ionizing photons that never reached the IGM along our line of sight.
This ``local reionization Soltan argument''
thus enables one to constrain 
the radiative efficiency of an \emph{individual} quasar via
its spectrum close to rest-frame
Ly$\alpha$.

\section{Constraints on the Radiative Efficiency of $z>7$ Quasars} \label{sec:constrain}

\begin{figure}[htb]
\begin{center}
\resizebox{8.5cm}{!}{\includegraphics[trim={1em 1em 1em 1em},clip]{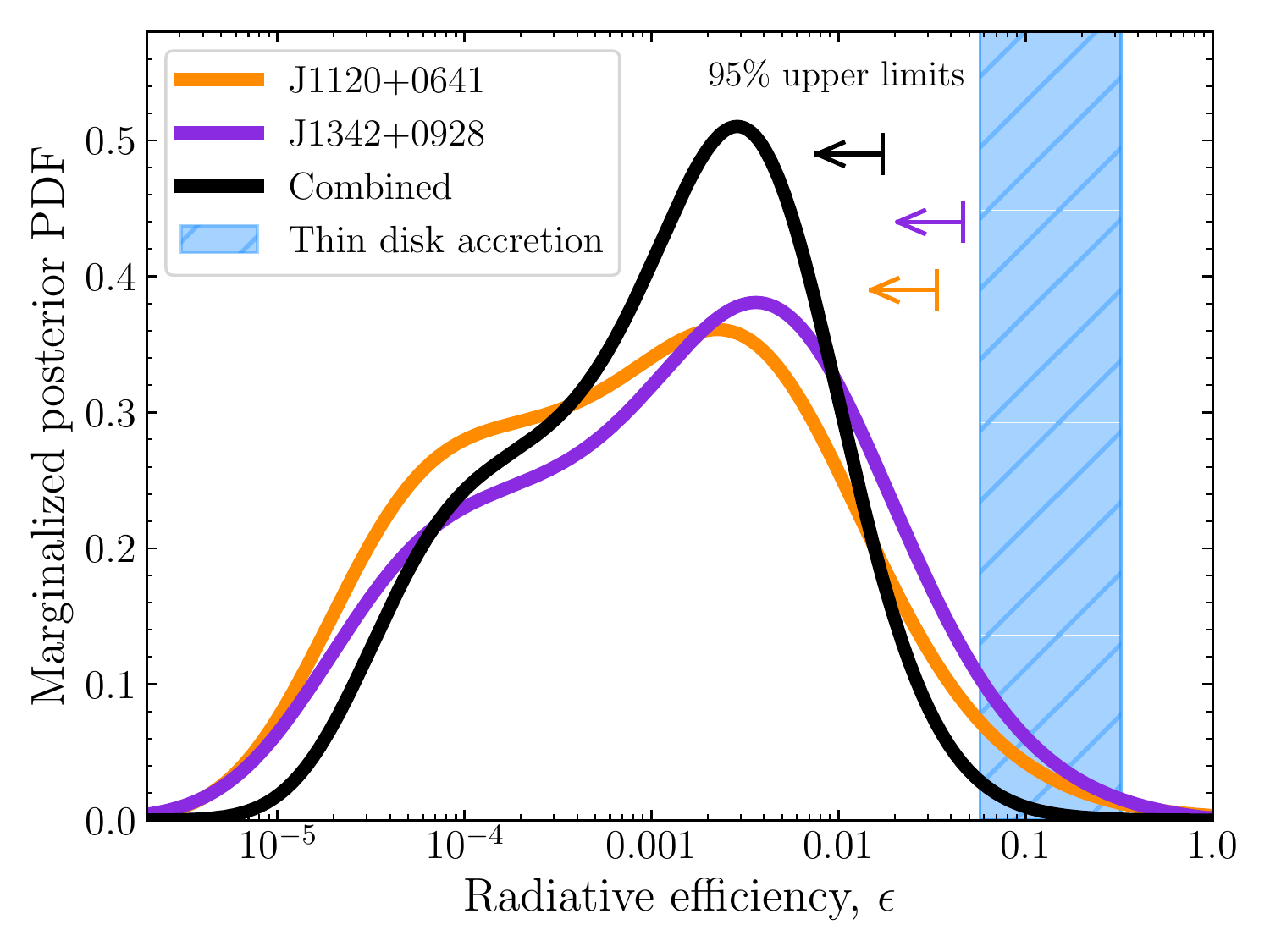}}
\end{center}
\caption{Marginalized posterior PDF of the radiative efficiency $\epsilon$. The orange and purple curves show the posterior PDFs of $\epsilon$ for J1120+0641 and J1342+0928, respectively, marginalized over obscuration and the systematic uncertainty in their black hole masses. The black curve shows the combined constraint under the assumption that both quasars have the same true radiative efficiency. Arrows indicate 95\% credibility upper limits on $\epsilon$. The blue shaded region shows the range of radiative efficiencies predicted for thin accretion disks in general relativity \citep{Thorne74}. }
\label{fig:epspost}
\end{figure}

\begin{figure*}[htb]
\begin{center}
\resizebox{15cm}{!}{\includegraphics[trim={1em 1em 1em 1em},clip]{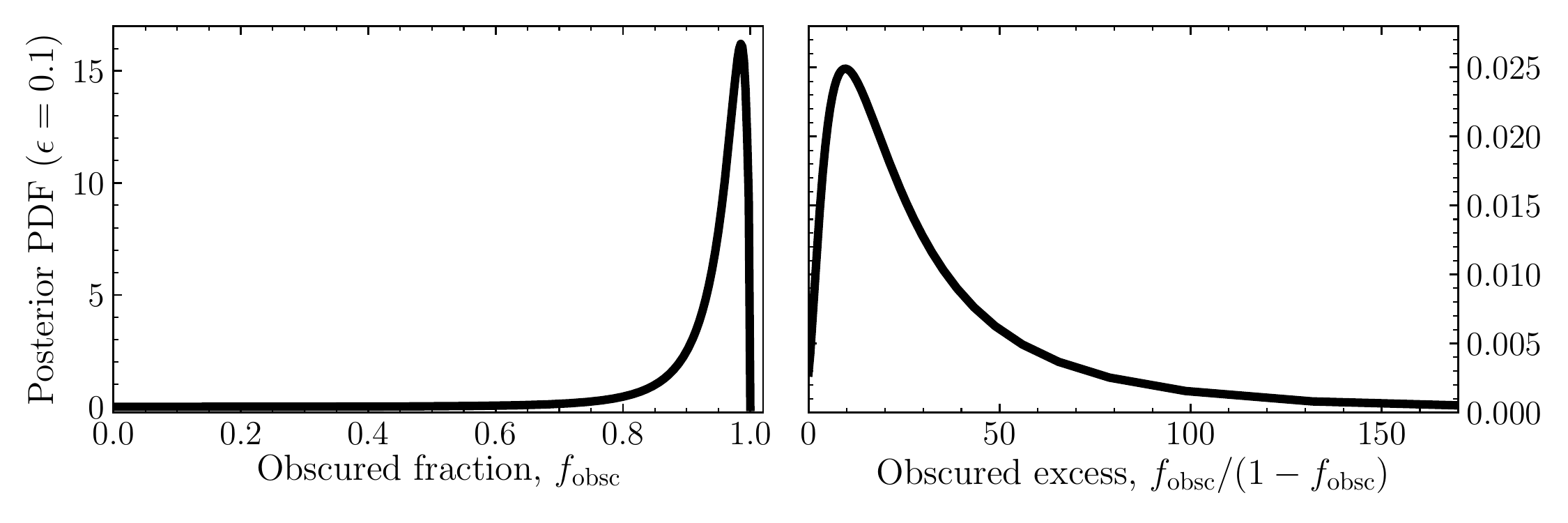}}
\end{center}
\caption{Constraints on the obscuration of $z>7$ quasars assuming $\epsilon=0.1$. Left: Posterior PDF for $f_{\rm obsc}$, combining the constraints from J1120+0641 and J1342+0928. Right: Combined posterior PDF for the relative number of obscured versus unobscured quasars, $f_{\rm obsc}/(1-f_{\rm obsc})$.}
\label{fig:fobsc}
\end{figure*}

We measured $N_{\rm ion}$
for the two highest
redshift quasars known, J1120+0641 and J1342+0928, by analyzing the
Ly$\alpha$ absorption in their rest-frame UV spectra in a very similar fashion to \citet{Davies18b}. 
The intrinsic, unabsorbed quasar spectrum close to rest-frame Ly$\alpha$
was estimated via a predictive principal
component analysis (PCA) approach from \citet{Davies18a}.
In Figure~\ref{fig:spectra} we show the two quasar spectra close to
rest-frame Ly$\alpha$ (grey and black curves) compared to their respective
PCA continuum models (blue curves). Both quasars show compelling evidence for an IGM damping wing
and truncated proximity zones, as previously shown by \citet{Davies18b}. 

We model reionization-epoch quasar spectra via a multi-scale approach following \citet{Davies18b} (see also Appendix~\ref{sec:method}). The large-scale topology of reionization around massive dark matter halos was computed in a (400 Mpc)$^3$ volume using a modified version of the \texttt{21cmFAST} code (\citealt{Mesinger11}; Davies \& Furlanetto, in prep.), and we stitched lines of sight through this ionization field onto skewers of baryon density fluctuations from a separate (100~Mpc$/h)^3$ \texttt{Nyx} hydrodynamical simulation \citep{Lukic15}. Finally, we performed 1D ionizing radiative transfer to model the ionization and heating of the IGM by the quasar \citep{Davies16,Davies19a}.
 
Through a Bayesian analysis on a grid of forward-modeled mock Ly$\alpha$ spectra from our simulations (Appendix~\ref{sec:method}), we jointly constrained the total number of
ionizing photons emitted by the quasars ($N_{\rm ion}$) and the
volume-averaged neutral fraction
of the IGM ($\langle x_{\rm HI} \rangle$). The mean Ly$\alpha$ absorption profiles of our best-fit
models and their 68\% scatter in the mock spectra are shown as the orange curves and shaded regions in Figure~\ref{fig:spectra}.  
The red curves in Figure~\ref{fig:spectra}
show alternative models where the IGM is fully neutral and $N_{\rm ion}$ for each quasar is instead determined via equation~(\ref{eqn:lrsa}),  
assuming $\epsilon=0.1$, $f_{\rm obsc}=0$, and $\Delta M_{\rm BH}=M_{\rm BH}$. 
These curves thus correspond to the maximum Ly$\alpha$ absorption in the standard
view of UV-luminous radiatively efficient SMBH growth.
The canonical radiative efficiency appears to be highly inconsistent with the data.

More quantitatively, in the bottom panels of Figure~\ref{fig:2dpost} we show the joint
posterior probability distribution functions (PDFs) for $N_{\rm ion}$
and $\langle x_{\rm HI} \rangle$ from our analysis of J1120+0641 (left) and
J1342+0928 (right). In the top panels of Figure~\ref{fig:2dpost} we show the marginalized posterior PDFs for
$N_{\rm ion}$.
Through the lens of equation~(\ref{eqn:lrsa}), we can view these marginalized posterior PDFs
 as constraints on the total
accreted black hole mass, indicated by the upper axes, where we assume $\epsilon=0.1$. 
The vertical lines show the
measured black holes masses for J1120+0641 and
J1342+0928, with shaded regions indicating their
systematic uncertainty.
For both quasars the inferred accreted mass is in
strong disagreement with the measured black hole mass,
or equivalently, a radiative efficiency much lower than 10\% is required to match the observations.

At face value, the results above indicate a serious inconsistency between
standard thinking about the radiative efficiency -- informed by general relativity,
accretion disk models, and the Soltan argument -- and our measurements
for these two reionization-epoch quasars. How can we reconcile the smaller than expected
number of ionizing photons emitted towards Earth with the observed
black hole masses? One possibility is that the bulk of the black hole
growth resulted in fewer ionizing photons
escaping into the IGM toward our line-of-sight due to obscuration by
gas and dust in the quasar host galaxy (e.g. \citealt{Hopkins05}). 
If the black holes grew appreciably during obscured phases, then
this is clearly degenerate with $N_{\rm ion}$
as indicated in equation~(\ref{eqn:lrsa}). 
Observations of similarly luminous quasars at
lower redshifts $z\gtrsim 2$ find that $\sim 50\%$ of them are
obscured \citep{Polletta08,Merloni14}, with some indication for increased obscuration at higher redshifts (\citealt{Vito18}, see also \citealt{Trebitsch19}).

To quantitatively constrain the radiative efficiency,
accounting for both the degeneracy with obscuration and
uncertainties in the
black hole masses, we re-map our 1D constraint on $N_{\rm
  ion}$ (i.e. the upper panels of Figure~\ref{fig:2dpost}) to a 3D space
of radiative efficiency, black hole mass, and the obscured fraction (see Appendix~\ref{sec:pdfmap}).
Marginalizing the 3D distributions over obscuration (assuming a uniform linear prior from 0 to 100\%) 
and black hole mass uncertainty (lognormal prior with $\sigma=0.4$~dex, \citealt{Shen13}) yields posterior PDFs for the radiative
efficiency as shown in Figure~\ref{fig:epspost}. The posterior median radiative efficiencies of J1120+0641 and J1342+0928
are 0.08\% and 0.1\%, respectively, and the canonical 10\%
is ruled out at greater than 98 per cent probability by each quasar. 
The combined posterior PDF for both quasars, assuming both quasars have the same true radiative efficiency, is shown by the black curve in
Figure~\ref{fig:epspost}, which is inconsistent with $\epsilon=0.1$ at 99.8 per cent
probability.

We can also assess what an assumed radiative efficiency of $10\%$ would imply for the obscuration of $z>7$ quasars. The left panel of Figure~\ref{fig:fobsc} shows the combined posterior PDF of $f_{\rm obsc}$ from both quasars assuming $\epsilon=0.1$, implying $f_{\rm obsc}>82\%$ at 95\% credibility. Such a high obscured fraction implies that there are many more similarly-luminous obscured quasars at $z>7$ which have not yet been identified. The right panel of Figure~\ref{fig:fobsc} shows the posterior PDF for the ratio of obscured to unobscured quasars, $f_{\rm obsc}/(1-f_{\rm obsc})$, which we computed from the $f_{\rm obsc}$ posterior PDF by a probability transformation. We constrain this ratio to be $25.7^{+49.6}_{-16.5}$ (posterior median and 68\% credible interval), with a 95\% credible lower limit of $4.4$.

\section{Discussion \& Conclusion} \label{sec:discus}

\begin{figure*}[htb]
\begin{center}
\resizebox{18cm}{!}{\includegraphics[trim={7em 3em 4em 0},clip]{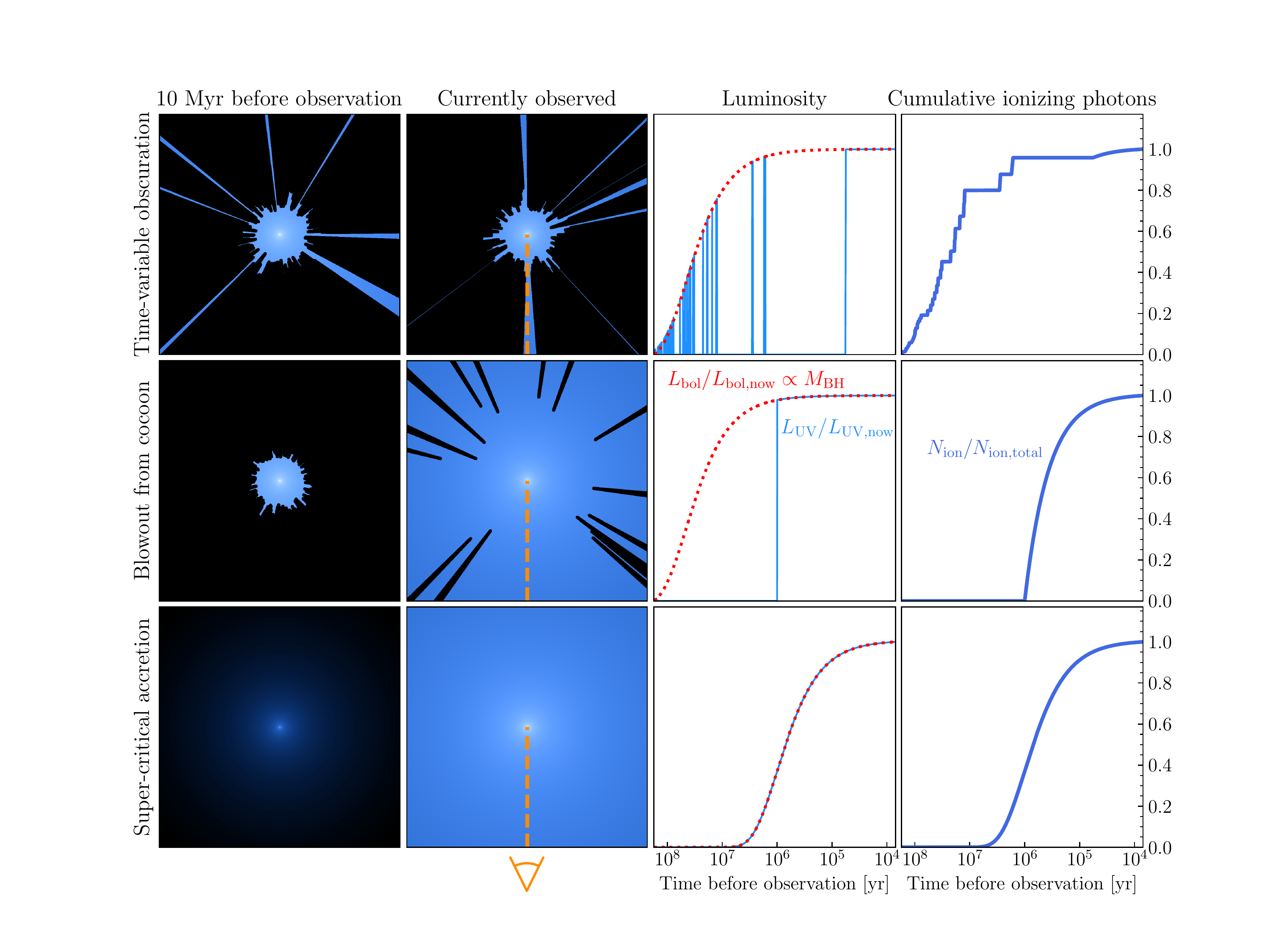}}
\end{center}
\caption{Possible solutions to the measured ionizing photon deficiency. The first two columns show schematic representations of the immediate environment of the $z>7$ quasars suggested by the measured deficit of ionizing photons. The blue regions are illuminated by the quasar while the black regions are not. The line of sight towards Earth is in the negative vertical direction as indicated by the orange dashed line. The leftmost column shows the quasar environment as it was 10 Myr ago, while the next column shows the quasar as it is observed today. The third column shows the UV (blue) and bolometric (red) luminosity history of the quasar, where we assume that the bolometric luminosity is proportional to the Eddington luminosity and thus its evolution indicates the growth of the black hole. The rightmost column shows the integrated number of ionizing photons that escaped into the IGM along our line of sight. The top and middle rows show two possible obscuration scenarios -- time-variable obscuration with a large covering fraction (top), or full obscuration followed by blowout (middle). The bottom row shows the case where the radiative efficiency is low, leading to rapid black hole growth and a much less luminous quasar 10 Myr ago.}
\label{fig:solutions}
\end{figure*}

If this radiatively inefficient mode of growth that we have uncovered
applies to quasars at later cosmic epochs, the Soltan argument implies previous
analyses have underestimated the present day SMBH mass density
by at least an order of magnitude.
Without invoking extra SMBH mass locally, the only solution is that
$z\gtrsim 7$ quasars grow or emit their radiation qualitatively differently
from their lower redshift counterparts. A few possibilities are
illustrated in Figure~\ref{fig:solutions}. It could be that $z>7$ quasar accretion disks are
actually radiatively efficient with $\epsilon \simeq 0.1$ but simply have
much more obscuration than inferred from studies of quasar demographics
at lower redshift (see also \citealt{Comastri15}). As our analysis only constrains the integrated
number of ionizing photons emitted in our direction, it is agnostic to
the exact nature of the obscuration.  It could have been highly time-variable
with obscured phases lasting $\gtrsim 10$ times longer than UV
luminous ones (top row of Figure~\ref{fig:solutions}), or the black hole could have grown while
fully enshrouded until a ``blowout'' event $\sim 1$ Myr ago when it
transitioned to a UV luminous phase (middle row of
Figure~\ref{fig:solutions}) \citep{Hopkins05}. Either of these obscuration scenarios
predicts many of comparably-luminous obscured quasars for every
UV luminous one at $z > 7$, as discussed above (Figure~\ref{fig:fobsc}). The obscured fraction would then have to
evolve very rapidly to avoid overproducing luminous obscured quasars at later
times. Nevertheless, if such a population exists at $z> 7$, future
mid-IR observations with JWST have the potential to uncover them.

Finally, let us not exclude the possibility that $z>7$ quasar accretion
disks are truly radiatively inefficient (bottom row of Figure~\ref{fig:solutions}).
This would
allow for rapid super-critical mass accretion rates with e-folding
timescales much shorter than 45 Myr without violating the
Eddington limit (equation~\ref{eqn:salp}), and has the appeal that it would easily explain
the existence of $\gtrsim 10^9$ SMBHs at early cosmic times $z > 7$ without
requiring overly massive seeds. 
This last scenario poses an intriguing
question: if the radiative efficiencies of the highest redshift quasars
are radically different from those at lower redshift, why do their
spectra appear nearly identical over eight decades in frequency \citep{Banados15b,Shen18,Nanni17}?
Similar to the original Soltan argument, the radiative efficiency that we have derived is a luminosity-weighted
average over the growth of the SMBH, which may differ from the efficiency of the currently
observed accretion flow.
Past phases of extremely super-critical accretion
cannot be ruled out, provided that they only occur at $z > 7$ -- in the same vein, however, 
neither can exotic formation scenarios, e.g. direct collapse to $10^9 M_\odot$, as long as they do not liberate UV photons. 
Future analyses of additional reionization-epoch quasars, combined with analogous measurements of the impact of luminous quasars on the IGM at lower redshifts \citep{Eilers18J1335,Schmidt18echo,Khrykin19,Davies19b}, will thus greatly improve our understanding of how SMBHs grew. 

\section*{Acknowledgements}
We thank Matthew McQuinn and Steven Furlanetto for comments on an early draft of this manuscript. FBD acknowledges support from the Space Telescope Science Institute, which is operated by AURA for NASA, through the grant HST-AR-15014.

\appendix

\section{Jointly Constraining the IGM \ion{H}{1} Fraction and $N_{\rm ion}$} \label{sec:method}

Here we summarize our methods for determining the intrinsic quasar continuum \citep{Davies18a} and Bayesian statistical analysis of reionization-epoch quasar transmission spectra \citep{Davies18b}. We refer the reader to \citet{Davies18a} and \citet{Davies18b} for further details on the methods employed.

\subsection{PCA Continuum Model}

We predict  the \citet{Mortlock11} Gemini/GNIRS spectrum of J1120+0641 and the \citet{Banados18} Magellan/FIRE+Gemini/GNIRS spectrum of J1342+0928 identically to \citet{Davies18a}. The intrinsic quasar continuum in the Ly$\alpha$ region (the ``blue side" of the spectrum, $\lambda_{\rm rest}=1180$--$1280$~{\AA}) was estimated via a principal component analysis (PCA) method built from a training set of $12764$ quasar spectra from SDSS/BOSS [refs] queried from the IGMSpec spectral database \citep{Prochaska17}. The red side of the quasar spectrum ($\lambda_{\rm rest}=1280$--$2850$~{\AA}) was fit to a linear combination of red-side basis spectra, and the best-fit coefficients were ``projected" to coefficients of blue-side basis spectra to predict the shape of the blue-side quasar spectrum.
While the systemic redshifts of J1120+0641 ($z=7.0851$; \citealt{Venemans16}) and J1342+0928 ($z=7.5413$; \citealt{Venemans17}) are very well known, the systemic redshifts of the training set quasars are considerably uncertain. We thus defined a standardized ``PCA redshift" frame by fitting the red-side coefficients simultaneously with a template redshift, and perform this same procedure when fitting the continua of the $z>7$ quasars. 

As shown in \citet{Davies18a}, the continuum uncertainty varies depending on the spectral properties of the quasar in question. We thus determined custom covariant uncertainty in the modeled continua by testing the procedure on $127$ SDSS/BOSS quasars with the most similar red-side spectra to each $z>7$ quasar. 

\subsection{Grid of Ly$\alpha$ Transmission Spectra}

Our numerical modeling of Ly$\alpha$ absorption in quasar spectra is identical to \citet{Davies18b}, as described in \S~\ref{sec:constrain}, however we re-computed the simulations from \citet{Davies18b} with a factor of five better sampling of quasar ages $t_{\rm q}$ to more carefully constrain $N_{\rm ion}\propto t_{\rm q}$.
We computed 2400 radiative transfer simulations for each IGM neutral fraction $\langle x_{\rm HI} \rangle$ in steps of $\Delta x=0.05$ from 0 to 1.0, with Ly$\alpha$ transmission spectra computed for $t_{\rm q}$ separated by $\Delta \log{t_{\rm q}}=0.1$ from $10^3$ to $10^8$ years. We later translated these quasar ages into $N_{\rm ion}$ by multiplying by the current ionizing photon output $\dot{N}_{\rm ion}$ for each quasar.

\subsection{Bayesian Statistical Method} \label{sec:stats}

Following \citet{Davies18b}, we performed a Bayesian statistical analysis by mapping out the likelihood function for summary statistics derived from forward-modeled mock data. We first bin the mock spectra to 500 km/s pixels, and fit 3 component Gaussian mixture models (GMM) to the flux distribution of each pixel for every pair of model parameter values $\theta=(\langle x_{\rm HI} \rangle,t_{\rm q})$ in our $21\times51$ model grid. We define a pseudo-likelihood,
\begin{equation}
\tilde{L}({\theta}) = \prod_i P_{{\rm GMM},i}(F_i|\theta),
\end{equation}
where $P_{{\rm GMM},i}(F_i|\theta)$ is the GMM of the $i$th pixel evaluated at its measured flux $F_i$ for model parameters $\theta$.

Treating the maximum pseudo-likelihood pair of parameter values $\theta_{\rm M\tilde{L}E}$ as a summary statistic to reduce the dimensionality of our data, we computed the posterior PDF of $\theta$ via Bayes' theorem,
\begin{equation}\label{eqn:post}
p(\theta|\theta_{\rm M\tilde{L}E}) = \frac{p(\theta_{\rm M\tilde{L}E}|\theta)p(\theta)}{p(\theta_{\rm M\tilde{L}E})},
\end{equation}
where $p(\theta|\theta_{\rm M\tilde{L}E})$ is the posterior PDF, $p(\theta_{\rm M\tilde{L}E}|\theta)$ is the likelihood function of $\theta_{\rm M\tilde{L}E}$,
 $p(\theta)$ is the prior, and $p(\theta_{\rm M\tilde{L}E})$ is the evidence. We assume a flat prior on our model grid, i.e. a flat linear prior on $\langle x_{\rm HI} \rangle$ and a flat logarithmic prior on $t_{\rm q}$; see \citet{Davies18b} for a discussion of the choice of these priors.
 
We explicitly compute the likelihood and evidence in equation~(\ref{eqn:post}) by measuring the distribution of $\theta_{\rm M\tilde{L}E}$ from forward-modeled mock observations on our coarse model grid of $\theta$. Each forward modeled spectrum consists of a random transmission spectrum from our set of 2400, a random draw from a multi-variate Gaussian approximation to the PCA continuum error (see \citealt{Davies18b}), and random spectral noise drawn from independent Gaussian distributions for each pixel according to the noise properties of the observed quasar spectrum.

\section{Deriving Radiative Efficiency Constraints} \label{sec:pdfmap}

Here we describe the method by which we convert our constraints on $N_{\rm ion}$ derived from the quasar spectra into constraints on the radiative efficiency $\epsilon$.
 The relationship between $N_{\rm ion}$ and $\epsilon$ described by equation~(\ref{eqn:lrsa}) involves two additional parameters, $f_{\rm obsc}$ and $\Delta M_{\rm BH}$, which are both uncertain. We thus recast our inference in terms of the set of parameters ${\bf a}\equiv\{\epsilon,f_{\rm obsc},\Delta M_{\rm BH}\}$ which are sufficient to determine $N_{\rm ion}$ through equation~(\ref{eqn:lrsa}). We assume $\Delta M_{\rm BH} = M_{\rm BH}$, which is a good approximation as long as $M_{\rm BH} \gg M_{\rm seed}$.

\begin{figure}[htb]
\begin{center}
\resizebox{18cm}{!}{\includegraphics[trim={0 0 0 0},clip]{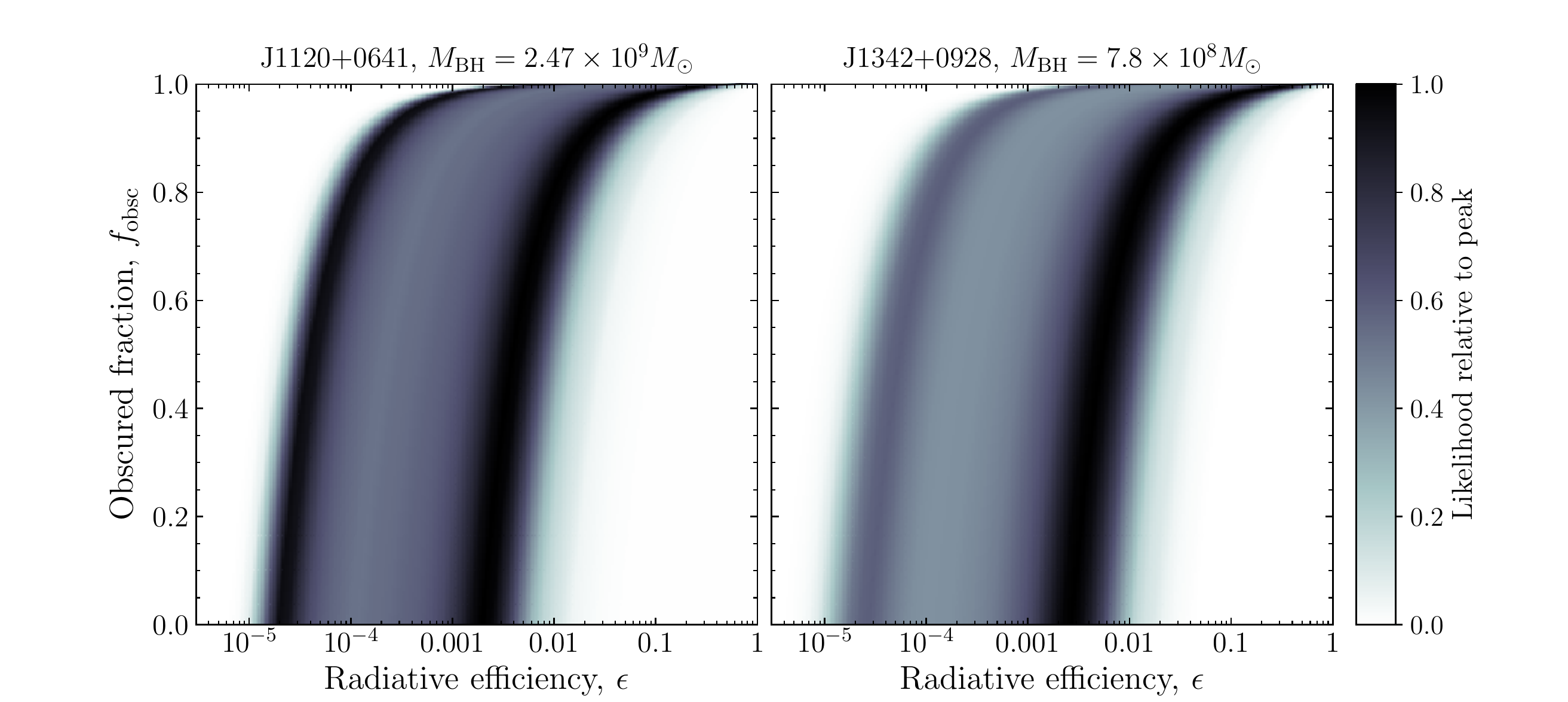}}
\end{center}
\caption{Slice through the 3D likelihood of $\epsilon$, $f_{\rm obsc}$, and $\Delta M_{\rm BH}$ for J1120+0641 (left) and J1342+0928 (right) at $\Delta M_{\rm BH}$ equal to their respective measured black hole masses.}
\label{fig:app_2dpdf_1}
\end{figure}

The likelihood function for the parameters in equation~(\ref{eqn:lrsa}), ${\bf a}\equiv\{\epsilon,f_{\rm obsc},\Delta M_{\rm BH}\}$, can be written as a marginalization over a joint likelihood of ${\bf a}$ and $N_{\rm ion}$,
\begin{equation}\label{eqn:like1}
p({\bf d}|{\bf a}) = \int p({\bf d}|{\bf a}, N_{\rm ion}) p(N_{\rm ion}|{\bf a}) dN_{\rm ion},
\end{equation}
where ${\bf d}$ represents the data. As described in the main text, the observed spectrum only depends on $N_{\rm ion}$, so $p({\bf d}|{\bf a}, N_{\rm ion})=p({\bf d}|N_{\rm ion})$. Additionally, we can write $p(N_{\rm ion}|{\bf a}) = \delta(N_{\rm ion}-\hat{N}_{\rm ion}({\bf a}))$, where $\hat{N}_{\rm ion}({\bf a})$ represents equation~(\ref{eqn:lrsa}) solved for $N_{\rm ion}$,
\begin{equation}\label{eqn:Nhat}
\hat{N}_{\rm ion}({\bf a}) = 2.3\times10^{72}\times \left(1-f_{\rm obsc}\right) \left(\frac{(1-\epsilon)/\epsilon}{9}\right)^{-1} \left(\frac{\Delta M_{\rm BH}}{10^9\,M_\odot}\right)
\end{equation}
 and $\delta$ is the Dirac delta function. Thus equation~(\ref{eqn:like1}) becomes
\begin{equation}\label{eqn:like2}
p({\bf d}|{\bf a}) = \int p({\bf d}|N_{\rm ion}) \delta(N_{\rm ion}-\hat{N}_{\rm ion}({\bf a})) dN_{\rm ion} = p({\bf d}|\hat{N}_{\rm ion}({\bf a})).
\end{equation}
In other words, the likelihood function of $\{\epsilon,f_{\rm obsc},\Delta M_{\rm BH}\}$ is equal to the likelihood function of $\hat{N}_{\rm ion}(\epsilon,f_{\rm UV},\Delta M_{\rm BH})$. Figure~\ref{fig:app_2dpdf_1} shows slices through this 3D likelihood for J1120+0641 and J1342+0928 at $\Delta M_{\rm BH}$ equal to their measured black hole masses of $2.47\times10^9\,M_\odot$ and $7.8\times10^8\,M_\odot$, respectively. With the likelihood for ${\bf a}$ in hand, we then marginalize over $\Delta M_{\rm BH}$ and $f_{\rm obsc}$ to recover a constraint on $\epsilon$ alone. 

\begin{figure}[htb]
\begin{center}
\resizebox{18cm}{!}{\includegraphics[trim={0 0 0 0},clip]{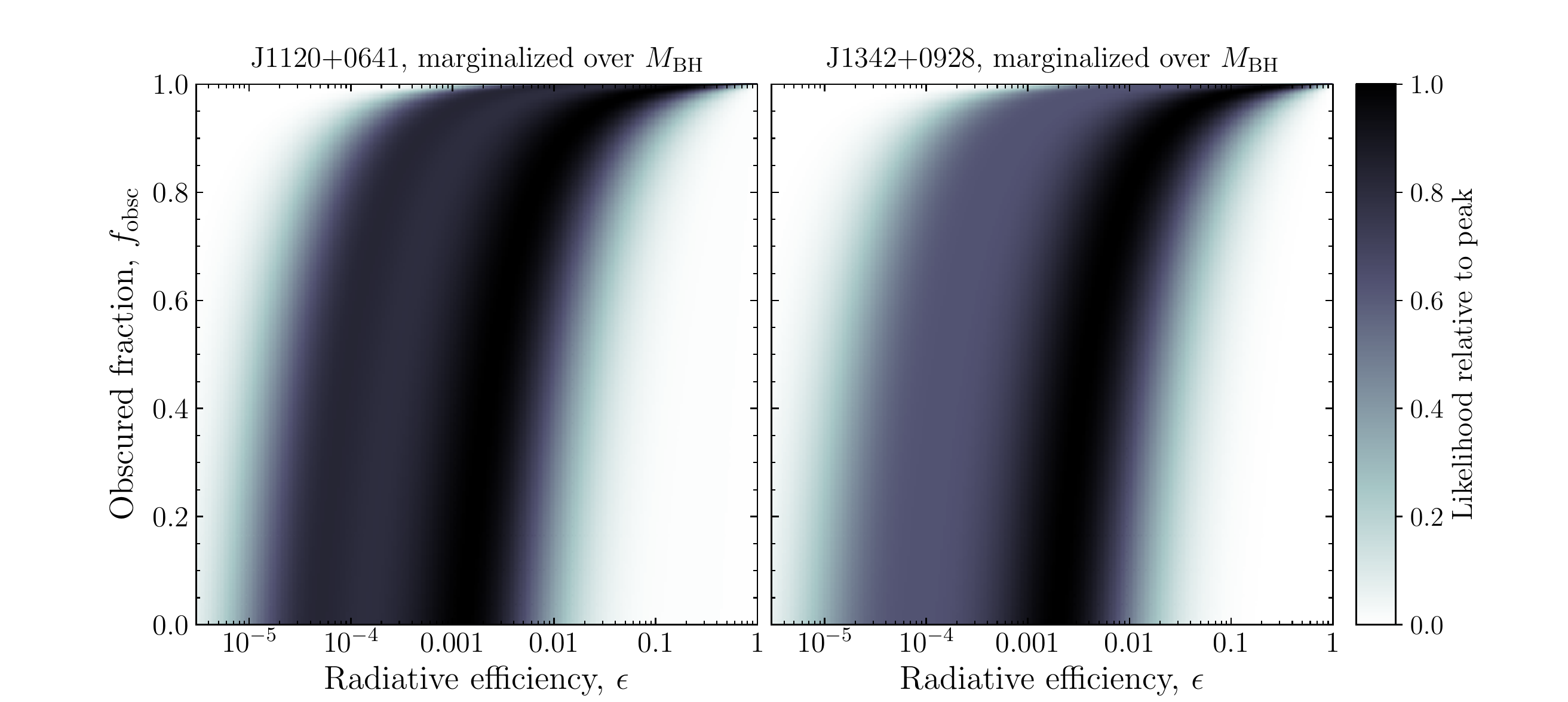}}
\end{center}
\caption{Similar to Figure~\ref{fig:app_2dpdf_1}, but now showing the joint likelihood of $\epsilon$ and $f_{\rm obsc}$ for J1120+0641 (left) and J1342+0928 (right) after marginalizing the 3D PDF from equation~(\ref{eqn:like2}) over a $\sigma=0.4$ dex lognormal systematic uncertainty in $M_{\rm BH}$.}
\label{fig:app_2dpdf_2}
\end{figure}

We marginalize over $\Delta M_{\rm BH}$ with a lognormal prior centered on the measured black hole mass with a 1$\sigma$ width of 0.4 dex \citep{Shen13}, resulting in the joint likelihood for $f_{\rm obsc}$ and $\epsilon$ shown in Figure~\ref{fig:app_2dpdf_2}.
We then marginalize over $f_{\rm obsc}$ with a uniform prior from 0 to 100\%. This prior on $f_{\rm obsc}$ reflects the fact that $\sim50\%$ of quasars with similar luminosity at lower redshift are obscured \citep{Polletta08,Merloni14} and that the evolution to $z>7$ is unknown. To subsequently derive the posterior PDF shown in Figure~\ref{fig:epspost}, we assume a log-uniform prior on $\epsilon$ from $10^{-7}$ to 1.

\bibliographystyle{apj}
 \newcommand{\noop}[1]{}

\end{document}